\documentclass[12pt]{article}
\usepackage{graphics,graphicx}
\usepackage{amssymb,epsfig,amsmath,euscript,array}

\makeatletter
\@addtoreset{equation}{section}
\makeatother

\newcounter{multieqs}

\newenvironment{pretty}{}{}

\begin{document}

\thispagestyle{empty}
\setcounter{page}{0}

\begin{flushright}
QMUL-PH-08-01
\end{flushright}

\vspace{20pt}

\begin{center}

{\Large \bf MHV One-Loop Amplitudes in Yang-Mills}
\\
\vspace{0.3cm}
{\Large \bf from Generalised  Unitarity}\\
\vspace{33pt}
{\bf  Andreas Brandhuber and Massimiliano Vincon}
\begin{pretty}\footnote{{\sffamily \{\tt a.brandhuber, m.vincon\}@qmul.ac.uk }}\end{pretty}

\vspace{2cm}

{\em Centre for Research in String Theory\\
Department of Physics\\
 Queen Mary, University of London\\
Mile End Road, London, E1 4NS\\
United Kingdom }\\
\vspace{45pt}
{\bf ABSTRACT}

\end{center}

\noindent
In this letter, we exploit  generalised unitarity in order to calculate the 
cut-constructible part of one-loop amplitudes 
in non-supersymmetric Yang-Mills theory. In particular, we rederive the $n$-gluon MHV amplitudes 
for both the adjacent and non-adjacent gluon helicity configurations from three- and 
four-particle cuts alone.
\newpage


\newpage
\tableofcontents

\newpage
\section{Introduction} \index{1}
\setcounter{footnote}{0}

It is common knowledge that the unitarity method, introduced in \cite{1,2} and further developed in \cite{3}, proved 
itself to be a powerful as well as elegant tool  for computing loop scattering amplitudes (see \cite{4} and references therein for a comprehensive review).
In fact, recent years have witnessed impressive achievements in the calculation of two- and higher-loop scattering 
amplitudes with much of the effort mostly focused on the maximally supersymmetric 
${\cal N}=4$ Yang-Mills theory (MSYM) \cite{5,6,7,8}. This is primarily due to the simplicity of 
the perturbative expansion in the 't Hooft (planar) limit of MYSM suggested by an intriguing duality 
that relates MSYM at strong coupling to weakly-coupled gravity on $AdS_{5} \times S^{5}$ \cite{9}. A short 
while ago, this duality was exploited as a different manner to compute amplitudes in MSYM
\cite{10} and in the case of four-gluon amplitudes 
agreement was found with an all-loop order ansatz put forward in \cite{6}.

In this letter, we focus on one-loop maximally helicity violating (MHV) amplitudes in pure Yang-Mills theory. These amplitudes are of particular interest as they constitute an example 
of one-loop $n$-point scattering amplitudes in QCD, where both external and internal particles are gluons. In pure Yang-Mills the $n$-gluon one-loop amplitudes may be decomposed as 
\begin{equation} \label{deco}
{\cal A}_{gluon}^{n} ={\cal A}_{{\cal N}=4}^{n}-4{\cal A}^{n}_{chiral, {\cal N}=1}+{\cal A}^{n}_{scalar} \, .
\end{equation}

Although each contribution of  (\ref{deco}) has been computed for the case of  MHV amplitudes using 
the unitarity method \cite{1,2},  the MHV diagram approach \cite{11,12,13,14} and, to some extent, generalised unitarity \cite{15,16}, 
an explicit double-check of the last term of (\ref{deco}), namely the contribution arising 
from a complex scalar particle running in the loop,  is still lacking for the case of MHV amplitudes with
non-adjacent negative-helicity gluons\footnote{So far that term has only been calculated using MHV diagrams in \cite{13}, while the 
special case of adjacent negative helicity gluons was first found in \cite{2}.}. 
As we felt obliged to do so, we aim in this letter to rederive the cut-constructible scalar contribution to the $n$-gluon  MHV amplitude by means of the generalised unitarity method \cite{17,18,3,15}.

At one loop, generalised unitarity instructs us to cut the amplitude into a product of up to four on-shell tree amplitudes and to replace the propagators connecting the sub-amplitudes by 
on-shell $\delta$-functions%
\footnote{Since the solutions of the momentum constraints can be complex in general we replace a cut propagator by 
$\delta(l_i^2)$ and not by $\delta^{(+)}(l_i^2)$. Also in the subsequent manipulations of the integrands
we allow the loop momenta to be complex.}, which put the internal particles on shell. 
When four propagators are cut (quadruple cut) the momentum integral is completely frozen and the
resulting product of four tree-amplitudes%
\footnote{To be more precise, in general the result is a weighted sum over the two complex solutions of the momentum constraints.} can be identified directly with coefficients of scalar
box functions \cite{15}. One route to obtain the coefficients for the remaining scalar triangle and bubble
functions is to use triple cuts and conventional two-particle cuts. An efficient
method to extract directly, individual coefficients of specific scalar integral functions using a convenient parametrisation for the cut momenta was presented recently in \cite{21b}.

For the extraction of triangle and bubble coefficients we want to follow a slightly different approach
\cite{16,20}. Here one considers the triple cut of a one-loop amplitude, which in general has contributions
from triangle and box functions. One can in principle subtract off the box contributions using quadruple
cuts but strictly speaking this is not needed. The three delta functions do not completely freeze the
loop integration, hence we simplify the integrand as much as possible using the three loop momentum
constraints where the loop momenta are allowed to take complex values. In the final step the cut integral
is lifted back up to a loop integral by replacing the on-shell delta functions by the corresponding
propagators. The result contains terms that have the correct cuts in the channel under consideration, and
possibly terms with cuts in other channels; the latter terms can be dropped. Considering all
possible cuts should then give the complete amplitude. An important comment is in order here. The
procedure outlined above also produces linear triangle integral functions (triangle integrals with
one loop momentum in the numerator), which, as is well known, can be written as linear combinations
of scalar triangle and bubble integrals. Therefore, this method can also produce bubble functions which
a priori would require the use of additional two-particle cuts. At this point we do not have a 
proof that two-particle cuts can be avoided for general amplitudes, but for the examples considered
in \cite{16,20} and in this letter this method produces the correct answers. The examples include the
Next-to-MHV one-loop amplitudes with adjacent negative helicity gluons considered in \cite{16},
all four-point one-loop amplitudes in pure Yang-Mills considered in \cite{20} and the MHV one-loop
amplitudes considered in this letter. Obviously, it would be 
interesting to study this observation in more detail.

In this paper we focus on the rederivation of the cut-constructible parts of MHV one-loop amplitudes
by considering a complex scalar running in the loop.
In the case that both negative helicity gluons are adjacent all quadruple cuts vanish and, hence, the
answer does not contain box functions. In the case that the negative helicity gluons are not adjacent
box functions do contribute and can be determined either directly using quadruple cuts (see \cite{16}) or with the triple cut method outlined above. As a consistency check we have
also considered the quadruple cuts in section 3.
Therefore, in the following discussion we will concentrate on the triple cuts, which in the case at hand
allow us to determine the full cut-constructible part of this class of amplitudes.
Explicitly, the non-vanishing triple cuts of the scalar loop contribution to the $n$-gluon MHV amplitude (see Figure 2) take the form:
\begin{eqnarray} \label{cut}
&&{\cal A}_{scalar}^{n}\Big|_{cut} = \\
&& \sum_{\pm}\int   d^{4}\ell_{1} \, d^{4}\ell_{2} \, d^{4}\ell_{3} \,\delta(\ell_{1}^{2}) \,
 \delta(\ell_{2}^{2}) \, \delta(\ell_{3}^{2}) \,\delta^{4}(\ell_{3}-\ell_{1}-Q) \delta^{4}(\ell_{1}-\ell_{2}-P){} \nonumber \\
&& \,\,\,\,\, \times {\cal A}_{tree}    (\ell_{1}, (m_2+1), \ldots,j^{-},\ldots,-\ell_{2}) {\cal A}_{tree}(\ell_{2},m_{1},-\ell_{3}) 
{\cal A}_{tree}(\ell_{3},\ldots,i^{-},\ldots, m_2, -\ell_{1}) \, , \nonumber
\end{eqnarray}
\\
\noindent
where  the allowed values of $m_{1}$ and $m_{2}$ are
\begin{equation}
j+1 \leq  m_{1} \leq  i-1, \,\,\,\,\,\,\,\,\,\, i+1 \leq m_{2} \leq  j-1 \, .
\end{equation}
The tree amplitudes entering the integrand involve two MHV amplitudes with two scalars
and one anti-MHV three-point amplitude with two scalars.
\noindent
The $\pm$ in (\ref{cut}) refers to the fact that we have a complex scalar running in the loop. Thus, there are two 
possible helicity configurations, each of which gives rise to the same integrand. 

On general grounds, four-dimensional cuts alone suffice to reconstruct the full amplitudes in  supersymmetric  theories at one loop \cite{1,2}. However, in theories not 
protected by supersymmetry,  there are additional rational terms which cannot be detected by  cuts, unless one decides  to work in 
$D=4 -2 \epsilon$  dimensions and keep higher orders in $\epsilon$, so that even rational terms develop discontinuities which can be 
detected by the unitarity method. An example of such an amplitude is the one-loop four-gluon 
$+++ \, +$ amplitude with a complex scalar running in the loop. This amplitude consists of purely rational terms and it was first computed in \cite{19} using a  technique based on the technology of four-dimensional heterotic string theory. It was subsequently confirmed and extended to the case of an arbitrary number
of positive helicity gluons in \cite{19a, chalmers} and to the case when one of the gluons has opposite helicity from the others \cite{19a}.
Furthermore, the $+++ \, +$ one-loop amplitude was recalculated 
in \cite{19b} by means of two-particle cuts in $D=4-2 \epsilon$ dimensions, in \cite{19c} where a relationship between one-loop MHV gluon amplitudes of QCD and those of ${\cal N}=4$ SYM was put forward and in \cite{20} using the generalised unitarity method in $D=4-2 \epsilon$ dimensions.
More recently, there has been a proposal \cite{21} in which it was argued that in a particular regularisation scheme certain Lorentz-violating counterterms provide these missing rational terms. 
We wish to make it clear  that in this letter we shall only work with unitarity cuts in $D=4$ dimensions, 
thus considering only the cut-constructible part of the $n$-gluon MHV amplitude.
Hence, all the (cut) loop momenta in this letter are kept in four dimensions until the amplitude
has been expressed as a linear combination of integral functions. Only at this stage the dimensional regularisation parameter $\epsilon$ is introduced to regularise the divergences of the integral
functions.

The cut-constructible part of the MHV one-loop amplitudes in pure Yang-Mills for the special case of adjacent negative helicity gluons has already been calculated in \cite{2} using unitarity  
whereas the general helicity configuration was dealt with in \cite{13} by means of the MHV diagram
method. Note that the rational parts of these amplitudes have been computed analytically in \cite{22,KosowerII} using the powerful method of on-shell recursion relations.
The purpose of this letter is to show how generalised unitarity correctly reproduces the 
cut-constructible parts of the
$n$-gluon amplitudes with less effort than conventional two-particle cuts or the MHV diagram method. We discuss the adjacent negative-helicity case in the next section and the general case in section 3. 
In section 4 we present our conclusions.

\setcounter{equation}{0}
\section{MHV one-loop amplitudes: adjacent negative-helicity gluons} \index{1}

In this section we  show how generalised unitarity may be used to compute the $n$-point MHV one-loop amplitude in pure Yang-Mills for the case of adjacent negative-helicity gluons.
\begin{figure}
\label{Figure1}
\begin{center}
\scalebox{0.7}{\includegraphics{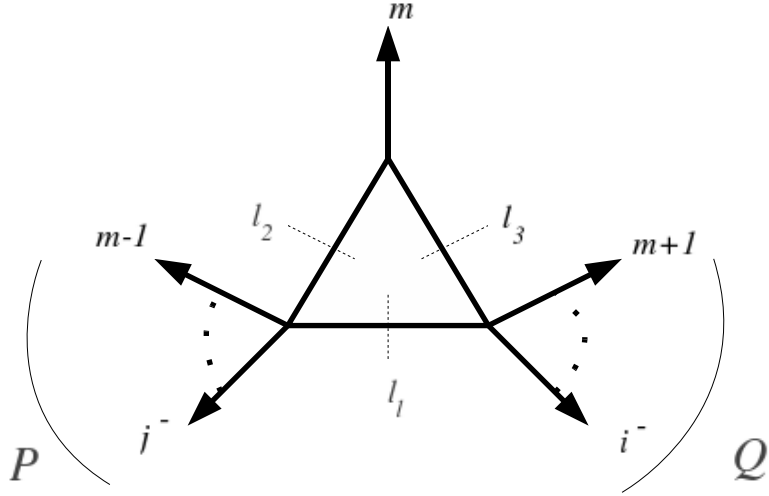}}
\end{center}
\caption{\em The three-particle cut diagram contributing to the $n$-gluon amplitude in the case of  adjacent 
negative-helicity gluons.}
\end{figure}

Let us consider the triple-cut diagram depicted in Figure 1, where we choose all momenta to be outgoing. 
There are two such diagrams, which are obtained by flipping all the internal helicities of the scalar particles running in the loop. Without loss of generality we set $i=1$ and $j=2$ throughout this section.
Note that in this case the range of $m$ is $3 \leq m \leq n$.
Furthermore, in the adjacent case all quadruple cuts vanish and, hence, 
no box functions appear in the amplitude.

The triple cut%
\footnote{For a short summary of conventions, see the Appendices.} 
of the $n$-point amplitude is obtained by sewing three tree-level amplitudes. 
Ignoring factors of $i$ and $2 \pi$, the product of the tree amplitudes appearing in the 
triple cut (\ref{cut}) is
\begin{eqnarray} \label{prod}
&&\frac{[ m \ell_2] [ m \ell_3]}{[ \ell_2 \ell_3]} \times
\frac{ \langle 1 \ell_1 \rangle^2 \langle 1 \ell_3 \rangle^2 }{\langle (m+1) (m+2)\rangle \ldots \langle n 1 \rangle \langle 1 \ell_1 \rangle \langle \ell_1 \ell_3 \rangle \langle \ell_3 (m+1) \rangle} \times \nonumber \\
&&\frac{ \langle 2 \ell_1 \rangle^2 \langle 2 \ell_2 \rangle^2 }{\langle 2 3\rangle \ldots \langle (m-2) (m-1) \rangle \langle (m-1) \ell_2 \rangle \langle \ell_2 \ell_1 \rangle \langle \ell_1 2 \rangle} \,\,\, .
\end{eqnarray}
Thus (\ref{cut}) together with (\ref{prod}) gives
\begin{eqnarray} \label{1}
&& {\cal A}^{n}_{scalar}\Big|_{cut}  \nonumber \\
&& =  2i A_{tree} \int \! \frac{d^{4} \ell_{2} \prod_{i=1}^3 \delta(l_i^2)}{(2 \pi)^{4}} 
\frac{\langle 2 \, \ell_{2} \rangle^{2} \, \langle \ell_{1} \, 2\rangle \, \langle 1 \, \ell_{1} \rangle \, \langle 1 \, \ell_{3} \rangle^{2} \, \langle (m-1) \, m \rangle \, \langle m (m+1) \rangle [m \, \ell_{2}] \, [\ell_{3} \, m]}
{\langle 1 \, 2 \rangle^{3}  \, \langle (m-1) \, \ell_{2} \rangle \, \langle \ell_{2} \, \ell_{1} \rangle \, 
\langle \ell_{3} \, (m+1) \rangle \, \langle \ell_{1} \, \ell_{3} \rangle \, [\ell_{2} \, \ell_{3}]}{} \nonumber \\
&&  = 2i \, A_{tree}  \int \! \frac{d^{4} \ell_{2}}{(2 \pi)^{4}} \, 
\frac{\langle 2 \,\big| \ell_{2} \big| m ]\, \langle 1\, m \rangle\, \langle 1 \, \big| P \ell_{2}\big|2\rangle\,
\langle 2 \big| P \ell_{2}\big| 1 \rangle\, \left[ 1 \,2 \right]^{3}}{2^{5}\,(1 \cdot 2)^{3}\,(\ell_{1} \cdot \ell_{2})^{2} \,  \ell_{1}^{2}\,\ell_{2}^{2}\,\ell_{3}^{2}}\Big|_{cut} \, ,
\end{eqnarray}
where in the second line of (\ref{1}) we have factored out the MHV tree level amplitude and cancelled
certain spinor brackets in the numerator and denominator of (\ref{prod}). In order to arrive at the last
line of (\ref{1})
we have used the fact that the holomorphic spinors of the momenta appearing in the anti-MHV three-point amplitude are proportional to each other, i.e. $\lambda_m \propto \lambda_{\ell_2} \propto \lambda_{\ell_3}$.
The factor of two accounts for the fact that we have already summed over the two possible internal helicities. Finally, the $\delta$-functions have been replaced by full propagators and the three-particle phase-space integral has been promoted to an unrestricted loop integral. The symbol $|_{cut}$ 
indicates that this replacement is only valid in the channel defined by a given triple-cut.

Let us clarify some notations. We define the general external momenta $k_{p}$ as $k_{p}:=p$. Also, we define 
\begin{equation}
P:=q_{j,m-1}, \;\;\;\;\;\; Q:=q_{m+1,i} \, ,
\end{equation}
where $q_{p_{i},p_{j}}:= \sum_{l=p_{i}}^{p_{j}}k_{l}$. We set $i=1$ and $j=2$ for the adjacent case.

Converting (\ref{1}) into Dirac traces yields the following integrand:
\begin{equation} \label{trace}
\frac{\textrm{tr}_{+}(\not \!1\!\not \!2\!\not \!P\!\not \!\ell_{2})\,\textrm{tr}_{+}(\not \!1\!\not \!2\!\not \!\ell_{2}\!\not \!m)\,\textrm{tr}_{+}
(\not \!2\!\not \!1\!\not \!\ell_{2}\!\not \!P)}
{2^{5}\,(1 \cdot 2)^{3}\,(\ell_{1} \cdot \ell_{2})^{2}} \, .
\end{equation}
Thus, the task  reduces to computing the three-index tensor integral
\begin{equation} \label{2}
{\cal I}^{\mu \nu \rho}(m,P,Q)=\int \! \frac{d^{4} \ell_{2}}{(2 \pi)^{4}} \, \frac{\ell_{2}^{\mu}\,\ell_{2}^{\nu}\,\ell_{2}^{\rho}}
{\ell_{1}^{2}\,\ell_{2}^{2}\,\ell_{3}^{2}} \, ,
\end{equation}
which may be done by  standard  Passarino-Veltman (PV) integral reduction \cite{23}.
Details of the calculation can be found in Appendix B.

The result of the PV reduction has to be inserted into (2.1). Doing so yields a 
series of terms of which, after some manipulations, only the following two remain:
\begin{eqnarray} \label{3}
A_{1} & = & -\frac{A_{tree}}{(t_{1}^{[2]})^{2}}\frac{1}{6}\frac{\left[I_{2}(P^{2})-I_{2}(Q^{2})\right]}{(Q^{2}-P^{2})^{2}}(1 \,2\,Q\,m)^{2}  \, , \\
A_{2} & = & \;\;\,\frac{A_{tree}}{(t_{1}^{[3]})^{3}}\frac{1}{3}\frac{\left[I_{2}(P^{2})-I_{2}(Q^{2})\right]}{(Q^{2}-P^{2})^{3}}(1\,2\,Q\,m)^{2}(1\,2\,m\,Q) \, ,
\end{eqnarray}
where $t_{i}^{[k]}:=(p_{i}+p_{i+1}+ \cdots +p_{i+k-1})^{2}$ are sums of color-adjacent momenta and the  $I_{2}$ functions are the scalar bubble functions as defined in Appendix A.
In obtaining (\ref{3}) and $(2.7)$, we made use of the fact that momentum conservation dictates that 
on the triple-cut \mbox{$(\ell_{1} \cdot \ell_{2})^{2}=4/P^{4}$} and $(m \cdot Q)= -(m \cdot P)=
-(1/2)(Q^{2}-P^{2})$. Also, in order to make the formulas more compact, we introduced the notation \mbox{$ (a_{1} \, a_{2} \, a_{3} \, a_{4}):=\textrm{tr}_{+}(\not \!a_{1}\!\not \!a_{2}\!\not \!a_{3}\!\not \!a_{4})$}, which we will use throughout the rest of the paper.

In $(2.6)$ and $(2.7)$ the combinations $\left[I_{2}(P^{2})-I_{2}(Q^{2})\right]/((Q^{2}-P^{2})^{(r)})$ appear, 
which are $\epsilon$-dependent  triangle functions expressed as differences of two bubble functions
(defined in Section A.1). For convenience we choose to write them as
\begin{equation} \label{4}
T^{(r)}_{\epsilon}(m,P,Q):=\frac{1}{\epsilon}\frac{(-P^{2})^{-\epsilon}-(-Q^{2})^{-\epsilon}}{(Q^{2}-P^{2})^{r}} \, ,
\end{equation}
where $r$ is a positive integer and the momenta on which $T^{(r)}$ depends satisfy $m+P+Q=0$.

As mentioned in the Introduction, we are working in $D=4$ dimensions so that we 
really should take the $\epsilon \rightarrow 0$ limit of (\ref{4}). 
For $P^2 \neq 0$ and $Q^2 \neq 0$ we define the finite, $\epsilon$-independent triangle function,\\
\begin{equation} \label{4bis}
T^{(r)}(m,P,Q):=\frac{\textrm{log}(Q^{2}/P^{2})}{(Q^{2}-P^{2})^{r}} \, .
\end{equation}

In the event of the vanishing of either of the kinematic invariants, (\ref{4}) gives rise to 
infrared-divergent terms since one of the numerator terms in (\ref{4}) vanishes. There are two possibilities:
\begin{itemize}
\item $P=k_{2}$ with $P^{2}=0$\, , \\
\item $Q=k_{1}$ with $Q^{2}=0$\, .
\end{itemize}

Finally, the amplitude takes the following form:
\begin{equation}
{\cal A}_{n}^{scalar}={\cal A}_{poles}+{\cal A}_{1}+{\cal A}_{2}\, ,
\end{equation}
where
\begin{eqnarray} \label{6}
{\cal A}_{poles} & = & -\frac{i}{6}{\cal A}_{tree}\frac{1}{\epsilon}\left[(-t_{2}^{[2]})^{-\epsilon}+(-t_{n}^{[2]})^{-\epsilon}\right]  \, ,\\
{\cal A}_{1} & = &   -\frac{2i}{6}{\cal A}_{tree}\frac{1}{(t_{1}^{[2]})^{2}} \sum_{m=4}^{n-1}\left[(1\,2\,P\,m)^{2}\right] T^{(2)}(m,P,Q) \nonumber \, , \\
{\cal A}_{2} & = &  -\frac{2i}{3}{\cal A}_{tree} \frac{1}{(t_{1}^{[2]})^{3}} \sum_{m=4}^{n-1} \left[(1\,2\,P\,m)^{2}(1\,2\,m\,P)\right]T^{(3)}(m,P,Q) \nonumber \, ,
\end{eqnarray}
where we used $t_{i}^{[k]}:=(p_{i}+p_{i+1}+ \cdots +p_{i+k-1})^{2}$ 
and the triangle functions introduced in (\ref{4bis}).

 Equation (\ref{6})\footnote{Notify that in the notation of \cite{2,13}
 $q_{m,1}=-P$. Also, we dropped an overall, $\epsilon$-dependent factor $c_\Gamma$ \cite{2} and did not make 
the symmetry properties of the amplitude under the exchange of the gluons $1\leftrightarrow2$ manifest in writing our result, thus explaining a factor of two compared to \cite{2,13}.}, 
which gives the cut-constructible part of the $n$-point one-loop scattering amplitudes with two adjacent gluons of negative helicity, agrees with the amplitudes
found in \cite{2} using conventional unitarity and with the amplitude 
found in \cite{13} using MHV diagrams.
\noindent

\setcounter{equation}{0}
\section{MHV one-loop amplitudes: non-adjacent negative-helicity gluons} \index{1}

The case in which the two negative-helicity gluons are non-adjacent is more involved. 
Fortunately, the calculation turns out to be more straightforward than expected, since some of
the algebraic manipulations involved can be related to manipulations appearing in the
MHV diagram calculation of the same amplitudes \cite{13}.

As in the adjacent case, our starting expression is (\ref{cut}). A direct, brute force calculation 
yields rather unpleasant four-tensor box integrals.
However, we do not follow this approach as it would spoil our goal to show the simplicity of 
the generalised unitarity method. Instead, by using momentum conservation arguments to eliminate 
$\ell_{3}$ from
(\ref{cut}), we arrive at a  more elegant and manageable expression for the amplitude given by
\begin{equation} \label{CUT}
{\cal A}_{scalar}^{n}\Big|_{cut}=-\frac{2i \, A_{tree}}{\langle i \,j \rangle^{4}}  \int \! \frac{d^{4}\ell_{2}}{(2 \pi)^{4}}
\frac{\langle j\, \ell_{1} \rangle^{2}\langle j\, \ell_{2}\rangle^{2} \langle i\, \ell_{1} \rangle^{2}
\langle i\, \ell_{2} \rangle^{2} \langle m_{2}\, (m_{2}+1) \rangle \langle (m_{1}\!-\!1)\, m_{1} \rangle[\ell_{2}\, m_{1}]}
{\ell_{1}^{2} \, \ell_{2}^{2} \, \ell_{3}^{2} \, \langle \ell_{1} \,(m_{2}\!+\!1) \rangle \langle (m_{1}\!-\!1)\, \ell_{2} \rangle \langle m_{2}\, \ell_{1} \rangle \langle \ell_{1}\, \ell_{2} \rangle^{2}} \Big|_{cut}\, ,
\end{equation}
where in deriving (\ref{CUT}) we made use of the fact that on the cut
\begin{eqnarray} \label{holo}
\lambda_{\ell_{2}} & = & \alpha \, \lambda_{m_{1}} \, , \\
\lambda_{\ell_{3}} & = & \beta \,  \lambda_{m_{1}} \nonumber \, , \\
\tilde{\lambda}_{\ell_{2}} & = & \frac{1}{\alpha}\tilde{\lambda}_{m_{1}}+
\frac{\beta}{\alpha}\tilde{\lambda}_{\ell_{3}} \nonumber \, , 
\end{eqnarray}
for some complex $\alpha$ and $\beta$  with $\lambda$ and $\tilde{\lambda}$ holomorphic and antiholomorphic spinors of negative and positive helicity 
respectively.

In order to reduce the hexagon integral (3.1) to a linear combination of box and triangle integrals, we notice that multiplying and dividing (\ref{CUT}) by $\langle \ell_{2} \,m_{1} \rangle$
 allows us to write the integrands\footnote{The reader might argue, in view of (\ref{holo}), 
that $\langle \ell_{2} m_{1} \rangle$ is zero which entails that we are effectively multiplying (3.1) by $\frac{0}{0}$. However, at this point we are off-shell as we have {\it uplifted} the cut integral to a Feynman integral by replacing on-shell $\delta$-functions by full Feynman propagators.}, after 
applying the Schouten identity twice, as a sum of four terms
\begin{equation}\label{CCC}
{\cal C}(m_{2}\!+\!1, \, m_{1})-{\cal C}(m_{2}\!+\!1, \, m_{1}\!-\!1)-{\cal C}(m_{2},m_{1})+{\cal C}(m_{2},m_{1}\!-\!1) \, ,
\end{equation}
where
\begin{equation}
{\cal C}(a,b):=\frac{\langle j \,\ell_{1} \rangle^{2}\langle j \,\ell_{2}\rangle \langle i\, \ell_{1} \rangle
\langle i\, \ell_{2} \rangle^{2}}
{\langle \ell_{1}\, \ell_{2} \rangle^{2} \langle i \, j \rangle^{4}}\cdot \frac{\langle i \,a \rangle \langle j \,b \rangle}
{\langle \ell_{1}\, a \rangle \langle \ell_{2}\, b \rangle} \, .
\end{equation}
\\
\begin{figure}
\label{figure2}
\begin{center}
\scalebox{0.7}{\includegraphics{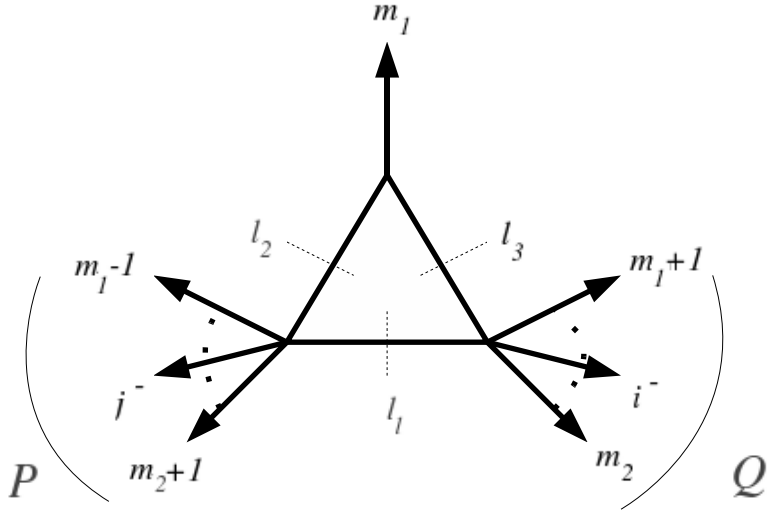}}
\end{center}
\caption{\em One of the two possible triple cut diagrams contributing to the $n$-gluon amplitude in the general case. The other triple cut diagram is obtained by swapping $i$ and $j$ through the replacements $m_{1}-1\rightarrow m_{1}$ and $m_{2} \leftrightarrow m_{1}$.}
\end{figure}
\\
\noindent
Therefore, we find 
\begin{equation} \label{Surprise}
{\cal A}_{scalar}^{n}\Big|_{cut}  =  2i \, A_{tree} 
\left[\int \! \frac{d^{4}\ell_{2}}{(2 \pi)^{4}}\frac{1}{\ell_{1}^{2} \, \ell_{2}^{2}}-\int \! \frac{d^{4}\ell_{2}}{(2 \pi)^{4}}\frac{1}{\ell_{1}^{2} \, \ell_{3}^{2}}\right]
\sum_{a,b} {\cal C}(a,b)
 \Big|_{cut}\, ,
\end{equation}
where the sum stands for the sum of four terms (with signs) in (\ref{CCC}).

One of the triple cuts contributing to the amplitude may be seen in Figure 2 where we defined $P:=q_{m_{2}+1,m_{1}-1}$ and $Q:=q_{m_{1}+1,m_{2}}$. Our choice for the momentum flow explains why we find the ${\cal C}$ coefficients with $a \leftrightarrow b$ compared to \cite{13}.

Although the calculation carried out in \cite{13} is conceptually different from the one we are performing here, we can nevertheless make use of formula (B.16) in that paper, which gives a rather convenient expression for ${\cal C}$:
\begin{eqnarray}
-{\cal C}(a,b) & = & \frac{(i\,j \,\ell_{1}\, \ell_{2})(i\,j \,\ell_{2} \,\ell_{1})(i\,j\, \ell_{1}\,a)(i\,j\,b\, \ell_{2})}{2^{8}(i \cdot j)^{4}(\ell_{1} \cdot \ell_{2})^{2}(\ell_{1}\cdot a)(\ell_{2} \cdot b)}{}  \\
&   = & \frac{1}{2^{8}(i \cdot j)^{4}}({\cal H}_{1}+\ldots+{\cal H}_{4}) \nonumber \, ,
\end{eqnarray}
where the ${\cal H}_i$ are given by
\begin{eqnarray}
{\cal H}_{1} & := & \,\,\,\,\, \frac{(i\,j\,b\,a)(i\,j \,\ell_{l} \,P)(i\,j\, P\, \ell_{1})(i\,j\, \ell_{1}\, a)}{(\ell_{1} \cdot \ell_{2})^{2}(a \cdot b)(\ell_{1} \cdot a)} {}  \\
& & \mbox{}- \frac{(i\,j\,b\,a)(i\,j\,P\, \ell_{2})(i\,j\, \ell_{2}\,P)(i\,j \,\ell_{2}\, b)}{(\ell_{1} \cdot \ell_{2} )^{2}(a \cdot b)(\ell_{2} \cdot b)} \nonumber \, , \\
\nonumber \\
{\cal H}_{2} & := & -\frac{(i\,j\,a\,b)(i\,j\,b\,a)(i\,j\, P\, \ell_{1})(i\,j\, \ell_{1}\, a)}{(\ell_{1} \cdot \ell_{2})(a \cdot b)^{2}(\ell_{1} \cdot a)} {}  \\
& & {}- \frac{(i\,j\,a\,b)(i\,j\,b\,a)(i\,j \,\ell_{2} \,P)(i\,j\, \ell_{2}\, b)}{(\ell_{1} \cdot \ell_{2} )(a \cdot b)^{2}(\ell_{2} \cdot b)}\nonumber  \, , \\
\nonumber \\
{\cal H}_{3} & := & -\frac{(i\,j\,a\,b)^{2}(i\,j\,b\,a)(i\,j\, \ell_{1}\, a)}{(a \cdot b)^{3}(\ell_{1} \cdot a)} {}  \\
& & {}+\frac{(i\,j\,a\,b)^{2}(i\,j\,b\,a)(ij \ell_{2} b)}{(a \cdot b)^{3}(\ell_{2} \cdot b)}\nonumber  \, , \\
\nonumber \\
{\cal H}_{4} & := & -\frac{(i\,j\,a\,b)^{2}(i\,j\,b\,a)^{2}(b\,P\, \ell_{1}\, a)}{4(a \cdot b)^{4}(\ell_{1} \cdot a)(\ell_{2} \cdot b)} \, .
\end{eqnarray}

Thus, we produce, in ascending order, linear box integrals and linear, two-tensor and three-tensor triangle integrals. We focus first on the triangle integral contributions.

Substituting for $a$ and $b$ in the expressions for $\cal{H}$ and keeping only those terms that actually contribute to the particular triple cut depicted in Figure 2 yields combinations of differences of traces. In order to express our result in a more compact fashion, we find it useful to define the following quantities:
\begin{eqnarray}
A_{m_{1}m_{2}}^{ij} & := & \frac{(i\,j\,m_{1}\,m_{2}\!+\!1)}{(m_{1} \cdot (m_{2}\!+\!1))}-\frac{(i\,j\,m_{1}\,m_{2})}{(m_{1} \cdot m_{2})} \, ,\\
S_{m_{1}m_{2}}^{ij} & := & \frac{(i\,j\,m_{1}\,m_{2}\!+\!1)(i\,j\, m_{2}\!+\!1\, m_{1})}{(m_{1} \cdot (m_{2}\!+\!1))^{2}}-\frac{(i\,j\,m_{1}\,m_{2})(i\,j \,m_{2}\,m_{1})}{(m_{1} \cdot m_{2})} \, , \\
I_{m_{1}m_{2}}^{ij} & := & \frac{(i\,j\,m_{1}\,m_{2}\!+\!1)(i\,j\, m_{2}\!+\!1 \,m_{1})^{2}}{(m_{1} \cdot (m_{2}+1))^{3}}-\frac{(i\,j\,m_{1}\,m_{2})(i\,j\, m_{2}\,m_{1})^{2}}
{(m_{1} \cdot m_{2})^{3}} \, ,
\end{eqnarray}
which exhibit the following symmetry properties
\begin{equation}
A_{m_{1}m_{2}}^{ij} = -A_{m_{1}m_{2}}^{ji}\, , \,\,\,\,\, S_{m_{1}m_{2}}^{ij} = S_{m_{1}m_{2}}^{ji} \, .
\end{equation}

The only integrals that survive from (\ref{Surprise}) are the ones with the correct triple cut, i.e. those integrals that have all three propagators that are cut in Figure 2. Hence, many of the  triangle integrals 
can be neglected%
\footnote{One consequence of these considerations is that the first integral on the right hand side of (\ref{Surprise}) can be ignored altogether.} and after the dust has settled
we are left with:
\begin{eqnarray}
{\cal H}_{1} & = & A_{m_{1}m_{2}}^{ij} \frac{(i\,j\,P\, \ell_{2})(i\,j\, \ell_{2}\,P)(i\,j\, \ell_{2}\,m_{1})}{2^{8}(i \cdot j)^{4}(\ell_{1} \cdot \ell_{2})^{2}(\ell_{2} \cdot m_{1})} \, , \label{eq1} \\
{\cal H}_{2} & = & S_{m_{1}m_{2}}^{ij} \frac{(i\,j\, \ell_{2}\,P)(i\,j\, \ell_{2}\,m_{1})}{2^{8}(i \cdot j)^{4}(\ell_{1} \cdot \ell_{2})(\ell_{2} \cdot m_{1})} \, ,  \label{eq2} \\
{\cal H}_{3} & = & I_{m_{1}m_{2}}^{ij} \frac{(i\,j\, \ell_{2}\,m_{1})}{2^{8}(i \cdot j)^{4}(\ell_{2} \cdot m_{1})} \, .
\label{eq3}
\end{eqnarray}

Before we present the complete amplitude, we wish to inspect the coefficient of the box function depicted in Figure 3 
and compare it with the results found in \cite{13} using MHV diagrams and in \cite{16}
using quadruple cuts. The crucial term in the function ${\cal C}(a,b)$ that enters the triple cut (\ref{Surprise}) of the amplitude and 
gives rise to a triple cut of a box function is:
\begin{equation}\label{quadcut}
-\frac{1}{2^8 (i \cdot j)^{4}}{\cal H}_{4} = \left[\frac{(i\,j\,m_{2}\,m_{1})^{2}(i\,j\, m_{1}\,m_{2})^{2}}{2^{8}(i \cdot j)^{4}(m_{2} \cdot m_{1})^{4}}\right]\frac{(m_{1}\,P \,\ell_{1}\,m_{2})}{4(l_{1} \cdot m_{2})(l_{2} \cdot m_{1})}\, ,
\end{equation}
\noindent
\begin{figure}
\label{figure3}
\begin{center} 
\scalebox{0.7}{\includegraphics{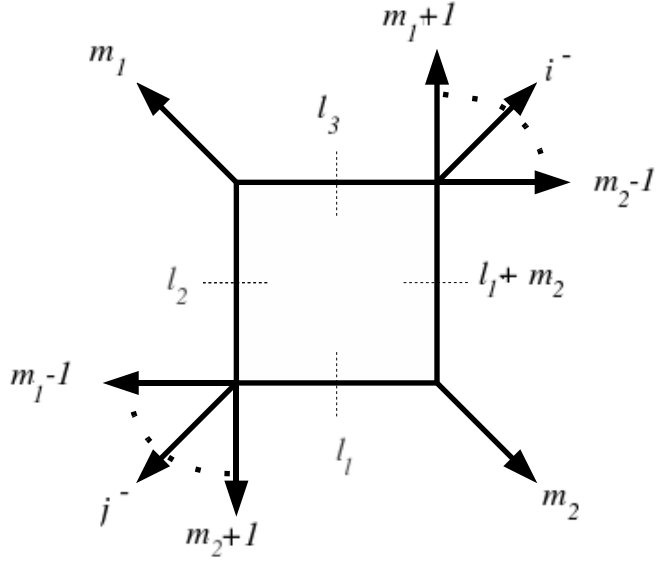}}
\end{center}
\caption{\em A box functions contributing to the $n$-gluon MHV amplitude in the general case.}
\end{figure}
which may be written more compactly as 
\begin{equation}\label{quadcut1}
-\frac{1}{2^8 (i \cdot j)^{4}}{\cal H}_{4} = \frac{1}{4}\left[b^{ij}_{m_{1}m_{2}}\right]^{2}\frac{(m_{1}\,P \,\ell_{1}\,m_{2})}{(l_{1} \cdot m_{2})(l_{2} \cdot m_{1})}\, ,
\end{equation}
in terms of the  coefficient of the  box integral function  appearing in the one-loop 
${\cal N}=1$ MHV amplitude with the same helicity configuration computed in \cite{2}
\begin{equation}
b^{ij}_{m_{1}m_{2}}:=-\frac{1}{8} \frac{(i\,j\,m_{2}\,m_{1})(i\,j\, m_{1}\,m_{2})}{(i \cdot j)^{2} \, (m_{1} \cdot m_{2})^{2}}  \, .
\end{equation}
Note that (\ref{quadcut1}) gives rise to a linear two-mass easy box integral whose PV reduction
has been performed in Appendix A.3. Inserting the result of this PV reduction into (\ref{quadcut1})
reproduces the correct coefficient of the box function. A brief comment is in order here. In the final result \cite{13} only the
finite part $B(s,t,P^2,Q^2)$ of the two-mass easy box function appears (as defined e.g. in eq. (4.7) of \cite{13}). We have checked that this is indeed the case and is due to the presence of scalar triangle
functions in the PV reduction of Appendix A.3 which precisely cancel the IR divergences of the scalar
box function $I_4[1]$ once all triple-cut channels are taken into account.

We can now present the complete result\footnote{We have already multiplied by a factor of 2 due to the two scalar helicity configurations running in the loop} for the one-loop $n$-gluon MHV amplitude (\ref{cut}) reconstructed using the generalised unitarity method:
\begin{eqnarray} \label{final}
{\cal A}^{n}_{scalar} & = & 2i \, {\cal A}_{tree} \left\{ \sum_{m_{1}={j+1}}^{i-1} 
\sum_{m_{2}=i}^{j-1} \frac{1}{2} [b^{ij}_{m_{1}m_{2}}]^{2} F
\left( t_{m_{1}}^{[m_{2}-m_{1}]},t_{m_{1}+1}^{[m_{2}-m_{1}-1]},P,Q \right)   \right.  \\
& & {}+\left( \frac{8}{3} \sum_{m_{1}=j+1}^{i-1}\sum_{m_{2}=i}^{j-1} 
\left[ {\cal A}^{ij}_{m_{1}m_{2}}T^{(3)}(m_{1},P,Q)+
(i \cdot j)\tilde{{\cal A}}^{ij}_{m_{1}m_{2}}T^{(2)}
(m_{1},P,Q)\right]{} \right. \nonumber \\
& & {} +2 \sum_{m_{1}={j+1}}^{i-1}
\sum_{m_{2}=i}^{j-1}\left[ {\cal S}^{ij}_{m_{1}m_{2}}T^{(2)}(m_{1},P,Q)-
{\cal I}^{ij}_{m_{1}m_{2}}T(m_{1},P,Q)\right]{} \nonumber \\
 & & {}  + (i \leftrightarrow j) \Bigg)  \Bigg\}  \nonumber \, ,
\end{eqnarray}
where we have introduced for convenience the following quantities:
\begin{eqnarray}
{\cal A}_{m_{1}m_{2}}^{ij}  & := & -2^{-8}(i \cdot j)^{-4}\,A_{m_{1}m_{2}}^{ij}\left[(i\,j\,m_{1}\,Q)(i\,j\,Q\,m_{1})^{2} \right] \, ,\\
\nonumber \\ 
\tilde{{\cal A}}_{m_{1}m_{2}}^{ij} & := & -2^{-8}(i \cdot j)^{-4}\,A_{m_{1}m_{2}}^{ij}\left[(i\,j\,Q\,m_{1})^{2}\right] \, ,\\
\nonumber \\
{\cal S}_{m_{1}m_{2}}^{ij} & := & \,\,\:\: 2^{-8}(i \cdot j)^{-4}\,S_{m_{1}m_{2}}^{ij}\left[(i\,j\,Q\,m_{1})^{2}\right] \, ,\\
\nonumber \\
{\cal I}_{m_{1}m_{2}}^{ij} & := & \,\,\:\: 2^{-8}(i \cdot j)^{-4}\,I_{m_{1}m_{2}}^{ij}\left[(i\,j\,Q\,m_{1})\right] \, .
\end{eqnarray}
The amplitude (3.21) agrees precisely with the result found in \cite{13}. Once again, in deriving (\ref{final}) we did not make use of the symmetry properties of the amplitude under exchange 
of the $i$-th and $j$-th gluon.

Similarly to the adjacent case, the infrared divergent terms may be extracted from the cases when either
$P^{2}$ or $Q^{2}$ vanishes (see Figure 2). The case $Q^{2}=0$ corresponds to $m_{1}=i\!-\!1$ and $m_{2}=i$, while $P^{2}=0$  corresponds to $m_{1}=j\!+\!1$ and $m_{2}=j\!-\!1$. Hence, 
\begin{eqnarray}
T^{(r)}(p,P,Q) & \rightarrow & (-)^{r}\frac{1}{\epsilon}\frac{(-t_{i-1}^{[2]})^{-\epsilon}}{(t_{i-1}^{[2]})^{r}}\, , \,\,\,\,Q^{2} \rightarrow 0 \, , \\
T^{(r)}(p,P,Q) & \rightarrow & \; -\frac{1}{\epsilon}\frac{(-t_{j}^{[2]})^{-\epsilon}}{(t_{j}^{[2]})^{r}}\, , \,\,\,\,\,\,\,\,\,\,\,\,P^{2} \rightarrow 0 \, .
\end{eqnarray}
Thus, we find the following infrared-divergent terms for $Q^{2}=0$:
\begin{eqnarray}
-\frac{1}{2 \, \epsilon} & \cdot & (-t_{i\!-\!1}^{[2]})^{-\epsilon}4(i \cdot j) \, \frac{(i \, j \, i\!-\!1 \, i\!+\!1)}{((i\!+\!1) \cdot (i\!-\!1))}{} \\
&  \cdot & \left[\frac{8}{3}(i \cdot j)^{2}-2 \, \frac{(i \, j \, i\!+\!1 \, i\!-\!1)}{((i\!+\!1) \cdot (i\!-\!1))( i \cdot j)}+\frac{(i \, j \, i\!+\!1 \,i\!-\!1)(i \, j \, i\!-\!1 \, i\!+\!1)}{((i\!+\!1) \cdot (i\!-\!1))^{2}}\right] \nonumber \, .
\end{eqnarray}
Similarly, we find for $P^{2}=0$ the following:
\begin{eqnarray}
-\frac{1}{2 \, \epsilon} & \cdot & (-t_{j}^{[2]})^{-\epsilon}4(i \cdot j) \, \frac{(i \, j \, j\!-\!1 \, j\!+\!1)}{((j\!+\!1) \cdot (j\!-\!1))}{} \\
&  \cdot & \left[\frac{8}{3}(i \cdot j)^{2}-2 \, \frac{(i \, j \, j\!+\!1 \, j\!-\!1)}{((j\!+\!1) \cdot (j\!-\!1))( i \cdot j)}+\frac{(i \, j \, j\!+\!1 \,j\!-\!1)(i \, j \, j\!-\!1 \, j\!+\!1)}{((j\!+\!1) \cdot (j\!-\!1))^{2}}\right] \nonumber \, .
\end{eqnarray}

\section{Conclusions}

We have shown how triple cuts correctly reproduce the cut-constructible part of the $n$-gluon one-loop 
MHV scattering amplitudes in pure Yang-Mills, both for the adjacent  and for the general case. 
An interesting observation of this calculation is that we did not have to make use of two particle
cuts. Of course, our result is consistent with two particle cuts since it agrees with the earlier calculation
of the same class of amplitudes in \cite{2} and \cite{13} using conventional unitarity and MHV diagram, respectively. This is in line with similar observations
made in \cite{16} and \cite{20} where certain classes of amplitudes where obtained from triple
cuts (and quadruple cuts) alone. The particular examples are the
Next-to-MHV one-loop amplitudes with adjacent negative helicity gluons considered in \cite{16} and
all four-point one-loop amplitudes in pure Yang-Mills considered in \cite{20}. 
Obviously, it would be interesting to investigate these observations further and understand
whether this works for general amplitudes.

A first, important step would be to gain knowledge of the one-loop $n$-gluon next-to maximally helicity 
violating amplitudes (NMHV), that is amplitudes with three negative helicities. While the purely gluonic 
6-\,, 7- and  $n$-point one-loop ${\cal N}=4$ NMHV amplitudes were computed in \cite{2,15,24,25} using generalised unitarity, 6- and  $n$-point one-loop amplitudes involving adjoint fermions and scalars in ${\cal N}=4$ gauge theory were found in \cite{26,27}. 
A different approach was employed in \cite{28} for the 7-gluon amplitudes in ${\cal N}=4$ NMHV, whereby the authors managed to exploit the holomorphic anomaly of unitarity cuts to reconstruct the amplitude by evaluating the action of a certain differential operator on the cut.
Furthermore, the holomorphic anomaly was also utilised in \cite{29} 
to compute the 6-point  one-loop ${\cal N}=1$ split-helicity NMHV amplitude, 
while the remaining 6-point one-loop ${\cal N}=1$ NMHV amplitudes were calculated in \cite{buchbinder}.
Generalised unitarity provided the $n$-gluon one-loop ${\cal N}=1$ NMHV amplitude in \cite{16} for the case that the three negative helicity gluons are adjacent. This latter amplitude has been
calculated in pure Yang-Mills in \cite{ita} using an iterative approach.
Finally, the coefficients of bubble and triangle integral 
functions for non-supersymmetric six-gluon amplitudes were computed in \cite{30}. 

\begin{figure}
\label{figure4}
\begin{center}
\scalebox{0.7}{\includegraphics{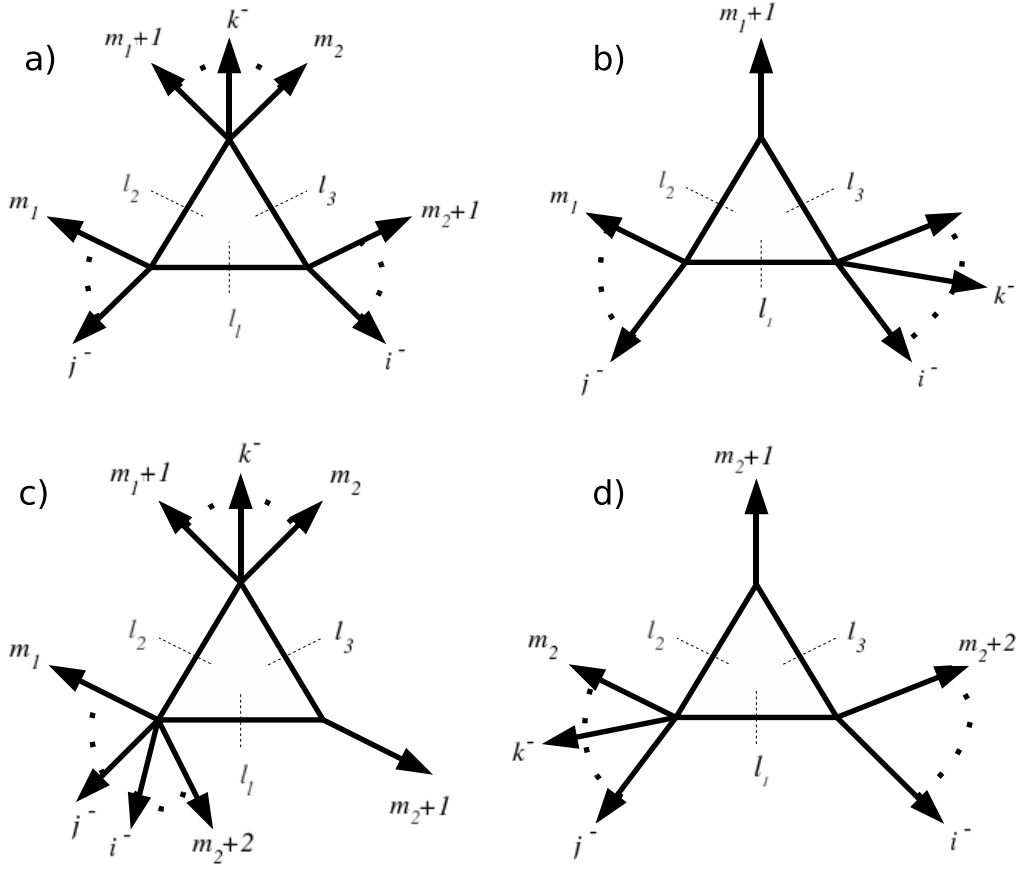}}
\end{center}
\caption{\em The triple-cut diagrams contributing to the $n$-gluon one-loop NMHV amplitude.}
\end{figure}

Let us conclude with some remarks on preliminary investigations of the NMHV case. We have started to investigate a particular class of non-supersymmetric NMHV amplitudes, namely 
 ${\cal A}^{n}_{scalar}(1^{+}, \ldots, i^{-},j^{-}, \ldots,k^{-},\ldots n^{+})$, i.e. amplitudes where the $i$-th and $j$-th 
negative helicity gluons are adjacent and the $k$-th one is in an arbitrary position. 
In order to tackle the problem, we start by 
identifying all  possible triple cuts contributing to the amplitude, which may be seen in Figure 4. The triple 
cut drawn in Figure 4a poses no new problems (we found structures similar to those appearing in
the calculation of the MHV amplitude we investigated in this letter). For the remaining triple cuts in Figure 4b, 4c and 4d  an additional difficulty arises, since the tree amplitudes appearing in the  triple cut (1.2) may  be NMHV. Thus, we cannot employ the Parke-Taylor formula for the standard MHV  tree amplitudes. In \cite{31}, it was shown how tree amplitudes in Yang-Mills theories may be derived by sewing together MHV vertices, suitably continued off-shell (CSW prescription), and connected by scalar bosonic propagators $1/p^{2}$ (see \cite{32} for a review). This novel diagrammatic approach stemmed from an insight which relates the perturbative expansion of ${\cal N}=4$ super 
Yang-Mills theory to D-instanton expansion in the topological B model on super twistor space 
$CP^{\rm 3|4}$ \cite{33}. By applying manipulations similar to those used in this letter, we mostly obtain three-tensor triangle integrals although some more complicated  three-tensor pentagon integrals still appear\footnote{In dealing with the NMHV tree amplitudes, we used the results of \cite{34,35,36}.}. In a straightforward application of the CSW rules  spurious poles arise and it 
is necessary to use improved formulas for the NMHV tree amplitudes \cite{35} that have only
physical poles. We plan to study this case in more detail in the future.

\section*{Acknowledgements}
It is a pleasure to thank Gabriele Travaglini for numerous discussions.
AB would like to thank STFC/PPARC for support under the grants PP/D507323/1 and PP/C50426X/1. 

\appendix
\setcounter{equation}{0}
\section{Tensor Integrals}
In this Appendix we define the one-loop integrals\footnote{We follow closely the conventions of \cite{2}. }
encountered in this paper, which were used in performing the PV  reductions. Furthermore, we
present formulas for the PV reductions of all tensor bubble, triangle and box integrals appearing
in our letter.
The more complicated three-tensor triangle integral  is dealt with separately in Appendix B.\\
\subsection{Bubble Integrals}
A general bubble integral is defined by
\begin{equation}
I_{2}[P(\ell^{\mu})]=-i (4 \pi)^{2}\, \int\frac{d^{4}\ell}{(2 \pi)^{4-2 \epsilon}}\frac{P(\ell^{\mu})}{\ell^{2}(\ell-K)^{2}} \, ,
\end{equation}
where $K$ is the total outgoing momentum at one side of the bubble and $P(\ell^{\mu})$ is some polynomial of the loop momentum $\ell^{\mu}$. Evaluation of the scalar bubble integral yields
\begin{equation} \label{bubble}
I_{2}[1]=r_{\Gamma}\frac{(-K^{2})^{-\epsilon}}{\epsilon(1-2\epsilon)}=r_{\Gamma}
\left[\left(\frac{1}{\epsilon}+2-\textrm{ln}(-K^{2})\right)+{\cal O}(\epsilon)\right],
\end{equation}
where
\begin{equation}
r_{\Gamma}=\frac{\Gamma(1+\epsilon)\Gamma^{2}(1-\epsilon)}{\Gamma(1-2\epsilon)}.
\end{equation}
Thus, we see that the difference of two scalar bubbles gives
rise to (\ref{4bis}) to ${\cal O}(\epsilon^{0})$.

The PV reductions of the linear and two-tensor bubble integrals are given by
\begin{eqnarray}
I_{2}[\ell^{\mu}] & = & -\frac{1}{2}I_{2}[1]K^{\mu} \, ,\\
I_{2}[\ell^{\mu} \ell^{\nu}] & = & I_2[1] \left( \frac{1}{3} K^{\mu}K^{\nu}-\frac{1}{12}K^{2}\eta^{\mu \nu} \right) \, .
\end{eqnarray}

\subsection{Triangle Integrals}
A general tensor triangle integral  is defined by
\begin{equation}
I_{3}[P(\ell^{\mu})]=i (4 \pi)^{2} \,\int\frac{d^{4}\ell}{(2 \pi)^{4-2 \epsilon}}\frac{P(\ell^{\mu})}
{\ell^{2}(\ell-K_{1})^{2}(\ell+K_{3})^{2}}\, ,
\end{equation}
where the $K_{i}$ are sums of the momenta $k_{i}$ of the external gluons at each vertex.
We find that the linear and two-tensor two-mass triangle integrals, with momentum assignments as in Figure 1, are given by
\begin{eqnarray}
I_{3}[\ell_{2}^{\mu}] & = & -T^{(1)}(m,P,Q) P^{\mu} + \ldots \, ,\\
I_{3}[\ell_{2}^{\mu}\ell_{2}^{\nu}] & = & \frac{1}{2}T^{(1)}(m,P,Q)P^{\mu}P^{\nu}-\frac{1}{2}P^{2}T^{(2)}(m,P,Q)\left(P^{\mu}m^{\nu}+P^{\nu}m^{\mu}\right) + \ldots \, .
\end{eqnarray}
The triangle functions $T^{(r)}(m,P,Q)$ have been defined in eq. (\ref{4bis}) and in the formulas above only those terms have been written down which survive after inserting the tensor integrals in the explicit cut expressions.
\noindent

\subsection{Box Integrals}
\noindent
A general tensor box integral is defined as
\begin{equation}
I_{4}[P(\ell^{\mu})]=-i (4 \pi)^{2} \,\int\frac{d^{4}\ell}{(2 \pi)^{4-2 \epsilon}}\frac{P(\ell^{\mu})}
{\ell^{2}(\ell-K_{1})^{2}(p-K_{1}-K_{2})^{2}(\ell+K_{4})^{2}}
\end{equation}
\noindent
For the linear box integral with momentum assignments as in Figure 3 we find
\begin{eqnarray}
I_{4}[\ell_{1}^{\mu}]& = & \frac{(m_{1} \cdot m_{2})\,P^{2}I_{4}[1]-(m_{1} \cdot P)\left[I_{3}+2 (m_{2}\cdot P) \, I_{4}[1]\right]}{2\left[(m_{1} \cdot m_{2}) \, P^{2}-2 \, (m_{2} \cdot P)(m_{1} \cdot P)\right]} P^{\mu}{}\\
\nonumber \\
& & {}+\frac{(m_{1} \cdot m_{2})\, P^{2}\left[I_{3}-(m_{2} \cdot P)I_{4}[1] \right]+(m_{1} \cdot P)(m_{2} \cdot P)\left[2 \, I_{4}[1] (m_{2}\cdot P)-I_{3}\right]}{2 (m_{1} \cdot m_{2}) \left[ (m_{1} \cdot m_{2}) \, P^{2}-2 \, (m_{2} \cdot P)(m_{1} \cdot P) \right]}m_{1}^{\mu} +\ldots  \nonumber \, , 
\end{eqnarray}
\noindent
where we are omitting a term proportional to  $m_{2}^\mu$ term since it drops out when inserted in (3.19). 
We refer the interested reader to the Appendices I and II of \cite{2} for a more complete discussion of bubble, triangle and box integrals.

\setcounter{equation}{0}
\section{Passarino-Veltman reduction}
In this section we carry out the PV reduction of the three-index tensor two-mass triangle integral, 
which enters in (\ref{2}) and (\ref{eq1})-(\ref{eq3}) and whose momentum assignments can be found in Figure 1:
\begin{equation} \label{A1}
{\cal I}^{\mu \nu \rho}(m,P,Q)=\int d^{4}\ell_{2} \frac{\ell_{2}^{\mu}\,\ell_{2}^{\nu}\,\ell_{2}^{\rho}}
{\ell_{1}^{2}\,\ell_{2}^{2}\,\ell_{3}^{2}} \, .
\end{equation}
The integral (\ref{A1}) may be decomposed as 
\begin{eqnarray} \label{A11}
{\cal I}^{\mu \nu \rho} & = & 
a(P^{\mu}P^{\nu}P^{\rho})+b(P^{\mu}m^{\nu}m^{\rho}+P^{\nu}m^{\mu}m^{\rho}+P^{\rho}m^{\nu}m^{\mu})+{}  \\
& & {} c(P^{\mu}P^{\nu}m^{\rho}+P^{\mu}P^{\rho}m^{\nu}+P^{\nu}P^{\rho}m^{\mu})+d(P^{\mu}\eta^{\rho \nu}+P^{\nu}\eta^{\mu \rho}+
P^{\rho}\eta^{\mu \nu})+{} \nonumber \\
& & {} e(m^{\mu}\eta^{\nu \rho}+m^{\nu}\eta^{\mu \rho}+m^{\rho}\eta^{\nu \mu})+f(m^{\mu}m^{\nu}m^{\rho}) \nonumber .
\end{eqnarray}
Taking contractions with all possible momenta then yields

\noindent
$\bullet P_{\mu} P_{\nu} P_{\rho}$
\begin{eqnarray} \label{A2}
 {\cal I}_{1}= \int \! \frac{(\ell_{2} \cdot P)^{3}}{\ell_{1}^{2}\,\ell_{2}^{2}\,\ell_{3}^{2}}& = & aP^{8}+3b[P^{2}(m \cdot P)^{2}]+
3c[(m \cdot P)P^{4}]+{}  \\
& & {}+3dP^{4}+3e[(m \cdot P)P^{2}]+f(m \cdot P)^{3} \nonumber \, ,
\end{eqnarray}
$\bullet P_{\mu} m_{\nu} m_{\rho}$
\begin{equation}
 {\cal I}_{2}=\int \! \frac{(\ell_{2} \cdot P)(m \cdot \ell_{2})^{2}}{\ell_{1}^{2}\,\ell_{2}^{2}\,\ell_{3}^{2}}=a[P^{2}(m \cdot P)^{2}]+
c(m \cdot P)^{3}+2d(m \cdot P)^{2} \, ,
\end{equation}
$\bullet P_{\mu} P_{\nu} m_{\rho}$
\begin{eqnarray}
 {\cal I}_{3}=\int \! \frac{(m \cdot \ell_{2})(\ell_{2} \cdot P)^{2}}{\ell_{1}^{2}\,\ell_{2}^{2}\,\ell_{3}^{2}} & = & a[(m \cdot P)P^{4}]+b(m \cdot P)^{3} 
 +2c[P^{2}(m \cdot P)^{2}]+{} \\
& & {}3d[P^{2}(m \cdot P)]+2e(m \cdot P)^{2} \nonumber \, ,
\end{eqnarray}
$\bullet P_{\mu}\eta_{\nu \rho}$
\begin{eqnarray}
 {\cal I}_{4}=\!\int \! \frac{(P \cdot \ell_{2})}{\ell_{1}^{2}\,\ell_{3}^{2}}\!&= & aP^{4}+2b[(m \cdot P)^{2}]+3c[P^{2}(m \cdot P)]+{} \\
& & {} 6dP^{2}+6e(m \cdot P) \nonumber \, ,
\end{eqnarray}
$\bullet m_{\mu}\eta_{\nu \rho}$
\begin{equation}
{\cal I}_{5}=\!\int \! \frac{(\ell_{2} \cdot m)}{\ell_{1}^{2}\,\ell_{3}^{2}}=a[(m \cdot P)P^{2}]+2c(m \cdot P)^{2}+6d(m \cdot P) \, ,
\end{equation}
$\bullet m_{\mu}m_{\nu}m_{\rho}$
\begin{equation}
{\cal I}_{6}=\!\int\!\frac{(\ell_{2} \cdot m)^{3}}{\ell_{1}^{2}\,\ell_{2}^{2}\,\ell_{3}^{2}}=a(m \cdot P)^{3} \, .
\end{equation}
The integrals take the following values:
\begin{eqnarray}
{\cal I}_{1} \!\!& = \!\! & \, -\frac{1}{2}(m \cdot P)^{2}I_{2}(Q^{2})-\frac{1}{8}P^{2}I_{3}{} \\
& & \mbox{} -\!\!\frac{1}{6}(P \cdot Q)^{2}I_{2}(Q^{2})+\frac{1}{24}Q^{2}P^{2}I_{2}(Q^{2}){} \nonumber\\
& & {} -\!\!\frac{1}{2}(m \cdot P)(P \cdot Q)I_{2}(Q^{2})+\frac{1}{4}(m \cdot P)I_{2}(Q^{2}){} \nonumber\\
& & {} +\!\!\frac{1}{8}P^{2}(P \cdot Q)I_{2}(Q^{2})-\frac{1}{8}P^{4}I_{2}(Q^{2}) \nonumber \, ,
\end{eqnarray}
\begin{equation}
{\cal I}_{2}  =\!-\frac{1}{6}(m \cdot Q)^{2}I_{2}(Q^{2})-\frac{1}{8}P^{2}(m \cdot Q)I_{2}(Q^{2})-
\frac{1}{8}P^{2}(m \cdot P)I_{2}(P^{2}) \nonumber \, ,
\end{equation}
\begin{eqnarray}
{\cal I}_{3}&=&\frac{1}{2}(m \cdot P)^{2}I_{2}(Q^{2})+\frac{1}{6}(P \cdot Q)^{2}I_{2}(Q^{2})
+\frac{1}{2}(m \cdot P)(P \cdot Q)I_{2}(Q^{2}){} \nonumber \\
& & {} -\frac{1}{24}Q^{2}P^{2}I_{2}(Q^{2})-\frac{1}{6}P^{4}I_{2}(P^{2})+\frac{1}{24}P^{4}I_{2}(P^{2}) \nonumber \, ,
\end{eqnarray}
\begin{equation}
{\cal I}_{4}=(m \cdot P)I_{2}(Q^{2})+\frac{1}{2}(P \cdot Q)I_{2}(Q^{2}) \nonumber \, ,
\end{equation}
\begin{equation}
{\cal I}_{5}\!=\!\frac{1}{2}(m \cdot Q)I_{2}(Q^{2}) \nonumber \, ,
\end{equation}
\begin{equation}
{\cal I}_{6}=\frac{1}{6}(m \cdot Q)^{2}I_{2}(Q^{2})-\frac{1}{6}(m \cdot P)^{2}I_{2}(P^{2}) \nonumber \, ,
\end{equation}
Finally, using Mathematica to carry out the algebraic manipulations, we retrieve the coefficients of 
the expansion (\ref{A11})
\begin{eqnarray}
a & = & \frac{I_{2}(Q^{2})-I_{2}(P^{2})}{3Q^{2}-3P^{2}} \, , \\ 
b & = & \frac{P^{4}(I_{2}(P^{2}-I_{2}(Q^{2})}{3(P^{2}-Q^{2})^{3}} \nonumber \, ,\\
c & = & \frac{P^{2}(I_{2}(Q^{2})-I_{2}(P^{2}))}{6(P^{2}-Q^{2})^{2}} \nonumber \, ,\\
d & = & \frac{Q^{2}I_{2}(Q^{2})-P^{2}I_{2}(P^{2})}{12(P^{2}-Q^{2})} \nonumber \, ,\\
e & = & \frac{(Q^{4}-2P^{2}Q^{2})I_{2}(Q^{2})+P^{4}I_{2}(P^{2})}{12(P^{2}-Q^{2})^{2}} \nonumber \, ,
\end{eqnarray}
where we chose not to write the $f$ coefficient as one can easily check that the $m_{\mu}m_{\nu}m_{\rho}$ term 
vanishes once inserted into the appropriate Dirac trace formulas appearing in our calculations. Incidentally, the $f$ coefficient is the only place where the $I_{3}$ scalar 
triangle function appears.

Thus, (\ref{A1}) takes the following form:
\begin{eqnarray}
\int d^{4} \ell_{2} \frac{\ell_{2}^{\mu}\,\ell_{2}^{\nu}\,\ell_{2}^{\rho}}{\ell_{1}^{2}\,\ell_{2}^{2}\,\ell_{3}^{2}} & = & \frac{I_{2}(Q^{2})-I_{2}(P^{2})}{3Q^{2}-3P^{2}}
(P^{\mu}P^{\nu}P^{\rho})+\frac{P^{4}(I_{2}(P^{2})-I_{2}(Q^{2}))}{3(P^{2}-Q^{2})^{3}}(P^{\mu}m^{\nu}m^{\rho}) \\
& & {} +\frac{P^{2}(I_{2}(Q^{2})-I_{2}(P^{2}))}{6(P^{2}-Q^{2})^{2}}(P^{\mu}P^{\nu}m^{\rho})+\frac{Q^{2}I_{2}(Q^{2})-P^{2}I_{2}(P^{2})}{12(P^{2}-Q^{2})}
(P^{\mu}\eta^{\nu \rho}) \nonumber\\
& & {} +\frac{(Q^{4}-2P^{2}Q^{2})I_{2}(Q^{2})+P^{4}I_{2}(P^{2})}{12(P^{2}-Q^{2})^{2}}(m^{\mu}\eta^{\nu \rho}) \nonumber \, .
\end{eqnarray}

\setcounter{equation}{0}
\section{Spinor Identities}
We list here some spinor identities. The Schouten identity is given by 
\begin{equation}
\langle i \,j \rangle \langle k \,l \rangle = \langle i\, l \rangle \langle k \,j \rangle + \langle i \,k \rangle \langle 
j\, l \rangle \, .
\end{equation} 
Other useful identities are
\begin{eqnarray}
 \left[ i\, j \right]\langle j\, i \rangle & = & \textrm{tr}_{+}(\not \!k_{i}\!\not \!k_{j})=2 (k_{i} \cdot k_{j}) \, , \\
 \left[i\,j \right]\langle j\,l  \rangle [l\,m]\langle m\,i \rangle  & = & \textrm{tr}_{+}(\not \!k_{i}\!\not \!k_{j}\!\not \!k_{l}\!\not \!k_{m}) \, .
\end{eqnarray}
In dealing with Dirac traces, we made use of the following identities:
\begin{eqnarray}
\textrm{tr}_{+}(\not \!k_{i}\!\not \!k_{j}\!\not \!k_{l}\!\not \!k_{m}) & = & \textrm{tr}_{+}(\not \!k_{m}\!\not \!k_{l}\!\not \!k_{j}\!\not \!k_{i}) \, = \, \textrm{tr}_{+}(\not \!k_{l}\!\not \!k_{m}\!\not \!k_{i}\!\not \!k_{j}) \, ,\\
\textrm{tr}_{+}(\not \!k_{i}\!\not \!k_{j}\!\not \!k_{l}\!\not \!k_{m}) & = & 4(k_{i} \cdot k_{j})(k_{l} \cdot k_{m})-\textrm{tr}_{+}(\not \!k_{j}\!\not \!k_{i}\!\not \!k_{l}\!\not \!k_{m}) \, .
\end{eqnarray}

\newpage
\addcontentsline{toc}{section}{References}
\bibliographystyle{unsrt}

\end{document}